\documentclass[prl,twocolumn,epsf,psfig]{revtex4}
\usepackage{graphicx}
\def\widetext{\end{twocolumn}}
\input epsf
\begin{document}
\title{Surface Tension in Unitary Fermi Gases with Population Imbalance}

\author{Theja N. De Silva$^{a,b}$,
Erich J. Mueller$^{a}$}
\affiliation{$^{a}$ LASSP, Cornell University, Ithaca, New York 14853, USA. \\
$^{b}$ Department of physics, University of Ruhuna, Matara, Sri
Lanka.}
\begin{abstract}
We study the effects of surface tension between normal and
superfluid regions of a trapped Fermi gas at unitarity. We find that
surface tension causes notable distortions in the shape of large
aspect ratio clouds. Including these distortions in our theories
resolves many of the apparent discrepancies among different
experiments and between theory and experiments.\end{abstract}

\maketitle

Experimentalists are now using dilute gases to controllably study
the properties of strongly interacting systems of superfluid
fermionic atoms \cite{biglistE, ketterle1,ketterle2,randy}.  Recent
experiments have examined the exotic circumstance where atoms with
two different hyperfine spins [denoted up and down] are placed in a
harmonic trap, but the number of spin-up atoms, $N_\uparrow$ is
greater than the number of down-spin atoms $N_\downarrow$
\cite{ketterle1,ketterle2,randy}.  Spin relaxation is negligible in
these experiments, so over the entire time of the experiment, the
system is constrained to have a fixed polarization
P=$(N_\uparrow-N_\downarrow)/(N_\uparrow+N_\downarrow)$.
Understanding the structure of (s-wave) superfluidity in this
polarized environment is an important endeavor with a long history
\cite{fflo,old1,old2,old3} and direct relevance to neutron stars,
thin-film superconductors, and color superconductivity.  In this
paper we use the concept of surface tension to quantitatively
explain controversial features seen in the density profiles of
strongly interacting trapped polarized Fermi gases \cite{randy,
ketterle1,ketterle2}.

The simplest theories of trapped Fermi gases
\cite{theja,chevy,new1,new2,ho} (most relying on local density
approximations [LDA] and assuming zero temperature) predict that the
atomic cloud phase separates into a central superfluid region, in
which the density of both spin species are equal, surrounded by a
polarized normal shell \cite{shells}. This basic structure was
observed in two separate experiments \cite{randy,
ketterle1,ketterle2},  however some experimental details are at odds
with these theoretical predictions. For $P>0.1$, the Rice
experiments \cite{randy} find a double peaked axial density
difference, $n_d^{(a)}(z)=\int dx\,dy\,[n_\uparrow({\bf
r})-n_\downarrow({\bf r})]$, where $n_{\uparrow/\downarrow}({\bf
r})$ is the density of up and down spin atoms. In a previous paper
\cite{theja}, we argued that this structure pointed to a breakdown
of the local density approximation, despite the fact that
dimensional arguments suggested that the LDA should work well.
Conversely, the results of the MIT experiments
\cite{ketterle1,ketterle2} are fully consistent with a local density
approximation, but show a polarization driven superfluid-normal
phase transition at $P\sim0.70$.  This phase transition was not seen
in the Rice experiments and is not found in most theories at
unitarity \cite{theja,chevy,new1,new2}. Here we show that surface
tension in the boundary between normal and superfluid regions
distorts the cloud in exactly the right way to account for the
unusual features seen at Rice.  We also show that surface tension
plays a much smaller role in the MIT experiments, where the atomic
clouds are larger and more spherical, and we are thus able to
account for the fact that the MIT experiment is consistent with the
local density approximation.  Finally, we show that for $P \gtrsim
0.7$, the Rice data shows a sudden drop in surface tension. Since
such a drop would be expected if the system underwent a
superfluid-normal phase transition, this observation may reconcile
the apparent differences in the experiments. We currently lack a
quantitative theory of the superfluid-normal phase transition at
unitarity.

In this letter we consider the unitary regime, where the scattering
length is infinite and the only lengthscale in the problem is the
interparticle spacing. Taking a two-shell structure, with a
superfluid core and a normal fluid shell, we model the free energy
of a trapped gas as
\begin{equation}\label{energy}
F=\int_Sd^3 r f_s[\mu(r),h] +\int_Nd^3 r f_n[\mu(r),h]
+\int_\partial d^2 r \sigma[\mu(r),h],
\end{equation}
where $\int_{S/N}$ represents the integral over the
superfluid/normal region, $\int_\partial$ corresponds to an integral
over the boundary, $f_{s/n}=-\int n^{(s)/(n)} d\mu$ represent the
free energy density of the superfluid/normal gas and  $\sigma$
represents the surface tension in the boundary.  The energy
densities are a function of the local chemical potentials
$\mu(r)=[\mu_\uparrow(r)+\mu_\downarrow(r)]/2=\mu_0-V(r),$ and
$h=[\mu_\uparrow(r)-\mu_\downarrow(r)]/2$, where
$V(r)=b_{\perp}\rho^2+b_{z}z^2=\frac{1}{2}m\omega^2(\lambda^2\rho^2+z^2)$
is the trapping potential, with $\lambda\approx50$ for the Rice
experiments and $\lambda\approx 5$ at MIT.  The shape of the
boundary, and the parameters $\mu_0$ and $h$, are determined by
minimizing eq.~(\ref{energy}) with respect to the boundary with the
constraint that $N_{\uparrow/\downarrow}=\int_S d^3r\
n^{(s)}_{\uparrow/\downarrow}+\int_N d^3r\
n^{(n)}_{\uparrow/\downarrow}$. This approach is a generalization of
one used by Chevy \cite{chevy}, where the boundary term was absent.
Universality allows us to write the  free energy density as
\begin{equation}
f_{s,n}(r)=\biggr(-\frac{2}{15 \pi^2}\biggr)
\biggr(\frac{2m}{\hbar^2}\biggr)^{3/2}\zeta_{s,n}
\mu_{s,n}(r)^{5/2},
\end{equation}
where $\zeta_{s}=1/(1+\beta)^{3/2}$, $\zeta_{n}=1/2$, and
$\beta\approx-0.545$ is a universal many-body parameter
\cite{parabeta}. The relevant chemical potentials are
$\mu_{s}(r)=\mu(r)$ and $\mu_{n}(r)=\mu_\uparrow(r)\equiv\mu(r)+h$.
The density of each spin component is
$n_{\uparrow,\downarrow}=-\partial f/\partial\mu$. The fact that the
particle spacing is the only length scale constrains the surface
tension to have the form $\sigma=(\hbar^2/2m) n_s^{4/3}(r)g(\delta
P/P)$, where $g$ is a function of $\delta P$, the pressure
discontinuity across the domain wall, and $P$, the pressure on the
superfluid side of the domain wall and $n_s$ is the density on the
superfluid side. Introducing a universal numerical parameter $\eta$,
we approximate $g$ by its value at zero pressure drop, $g(0)=\eta$,
based on estimating that $\delta P/P< 1.8 \times 10^{-3}$
~\cite{note1}.

We determine $\eta$ in two ways.  First, as detailed in the
appendix, a mean-field theory gradient expansion yields $\eta\approx
0.9 \times 10^{-3}$. Second, we use a fitting scheme where we
minimize Eq. (~\ref{energy}) for a series of candidate $\eta$'s. We
find that $\eta=1.0 \times 10^{-3}$ matches the Rice group's
experimental data for the axial density difference at P=0.53. Given
the uncontrolled nature of the mean field approximation, we believe
that the similarity of the two results is purely coincidental. We
use $\eta=1.0 \times 10^{-3}$ for all of our predictions.

To simplify the minimization of Eq. (\ref{energy}) with respect to
the boundary, we make the ansatz that the boundary is an ellipsoid
with semi major and minor axes $\overline{\rho}$ and $\overline{z}$.
Within this ansatz we analytically calculate the free energy [for
brevity we omit the expressions]. We minimize this expression with
respect to the parameters $\overline{\rho}$ and $\overline{z}$.

\begin{figure}
\includegraphics[width=\columnwidth]{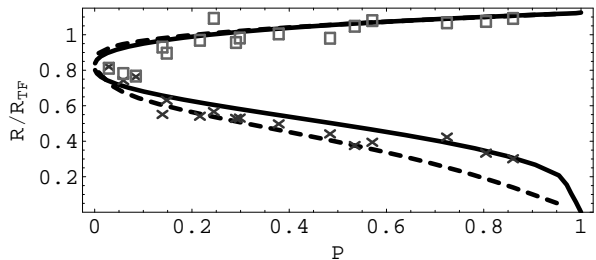}
\caption{ TOP: Comparison of the minority and majority components
radii with the Rice experiments~\cite{randy}. Squares and crosses
are the majority and minority components observed in \cite{randy},
determined by looking for where the density of each component
vanishes. The solid line is our theoretical prediction of the radii
with zero surface tension, while the dashed line includes finite
surface tension. The radii are normalized with a non-interacting
Thomas-Fermi radius, $R_{TF}=\sqrt{\epsilon_{f}/b_z)}$, where
$\epsilon_{f}= \hbar\overline{\omega}(6N)^{1/3}$ with average trap
frequency $\overline{\omega}=(\omega_{\perp}^{2}\omega_{z})^{1/3}$
and $N=(N_\uparrow+N_\downarrow)/2.$ Note, both the normalization
and the fitting procedure differ from the one used in figure 3 of
\cite{randy}. }\label{radii}
\end{figure}

\begin{figure}
\includegraphics[width=\columnwidth]{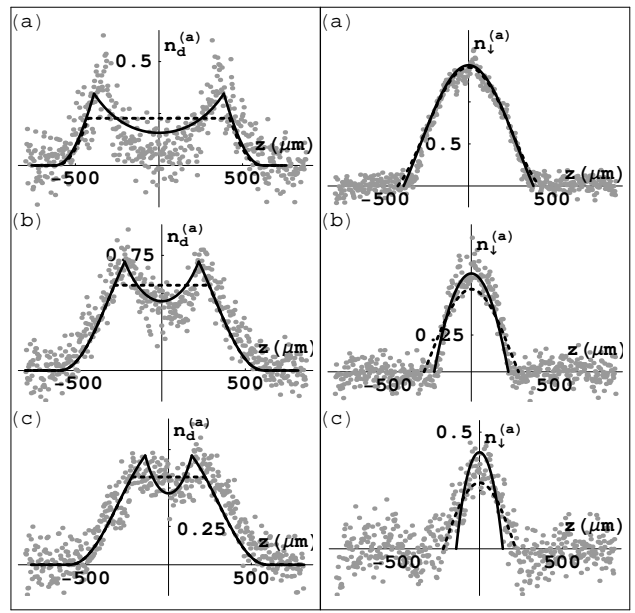}
\caption{ LEFT: Axial density difference $n_d^{(a)}(z)=2\pi\,\int
d\rho\,\rho\, [n_{\uparrow}(z,\rho)-n_{\downarrow}(z,\rho)]$ of zero
temperature harmonically trapped unitary Fermi gas in units of
$[10^6 cm^{-1}]$. Figures (a), (b), and (c) represent polarization
$P= 0.14$, 0.53, and 0.72 respectively. The grey points are the
experimental data from reference ~\cite{randy}. The $P=0.14$ and
$P=0.53$ data previously appeared in figure 2 of \cite{randy}. The
$P=0.72$ data corresponds to one of the points in figure 3 of ref
\cite{randy}. The dashed line is the zero surface tension density,
while solid line is the finite surface tension density using
$\eta=1.0 \times 10^{-3}$. RIGHT: Comparison of the axial density of
minority component $n_\downarrow^{(a)}(z)=2\pi\,\int d\rho\,\rho\,
n_{\downarrow}(z,\rho)$.}\label{difdw}
\end{figure}
To estimate the distortions, one expands Eq. (\ref{energy}) for
small distortions: $\overline{\rho}=\rho_0(1+\delta_{\rho})$, and
$\overline{z}=z_0(1+\delta_{z})$, where $\rho_0$ and $z_0$ are the
lengths of the axes in the absence of surface tension. We take
$\delta_{\rho}$ and $\delta_z$ to be order of $\delta$. Dimensional
analysis gives, $\delta F/(\hbar^2/2m) \sim A\ \eta n^{4/3}z_0
\rho_0 \delta+B\ n^{5/3}z_0 \rho_0^2 \delta^2$, where $A$ and $B$
are constants. Assuming that $\rho_0$ scales with the radial
Thomas-Fermi radius, the size of the distortion is then $\delta \sim
1/(\rho_0n^{1/3}) \sim (\lambda/N)^{1/3}$.

FIG. ~\ref{radii} shows the calculated axial radii as a function of
polarization. We compare our predictions to radii that we extract
from the data used in FIG. 3 of ref. \cite{randy}. We extract the
radii by fitting the wings of the axial density distributions to a
piecewise linear function of the form $n(z)=w\left(1-z/R\right)
\theta (R-z)$, where $\theta(x)=1$ for $x>0$ and $0$ otherwise, and
$w$ and $R$ are fitting parameters. This fitting procedure lets us
accurately determine the edge of the cloud, while the radii
extracted in \cite{randy} correspond to an average radius, and are
systematically larger those extracted by our method. We see that for
$0.1 \lesssim P \lesssim 0.7$ the experimental data is in excellent
agreement with the finite surface tension theory. For $P \gtrsim
0.7$, the data used for FIG. \ref{radii} appears to be inconsistent
with $\eta=1.0 \times 10^{-3}$. We speculate that the deviation may
be due to the superfluid-normal phase transition observed at
MIT~\cite{ketterle1,ketterle2}. We caution, however, that there are
large fluctuations in the radii seen in ref. \cite{randy} especially
at large P. More work is needed before definitive statements can be
made. The disagreement of the radii below $P_c \approx 0.1$ is
probably attributable to finite temperature effects \cite{pcom}.

\begin{figure}
\includegraphics[width=\columnwidth]{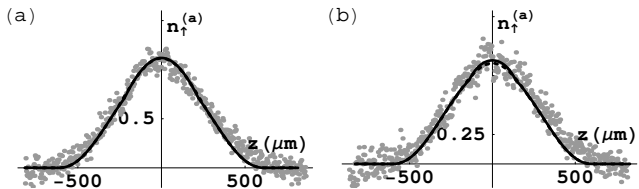}
\caption{ Comparison of the axial majority component density
$n_\uparrow^{(a)}(z)=2\pi\,\int d\rho\,\rho\, n_{\uparrow}(z,\rho)$
in units of $[10^6 cm^{-1}]$ with experimental data of reference
\cite{randy}. Figure (a) and (b) represent polarization $P=0.53$ and
$P=0.72$ respectively. Symbols carry the same meanings as in Fig
\ref{difdw}. Notice that the solid line and the dashed line
coincide, indicating surface tension has no effect on the majority
densities.}\label{up}
\end{figure}

In FIG. ~\ref{difdw} and \ref{up} we compare our predicted axial
density profiles with representative data from ref. \cite{randy}. As
demonstrate in the left panel of FIG.~\ref{difdw}, the finite
surface tension theory captures the observed double peak structure
in axial density difference for $P<0.7$.  The only free parameter in
this calculation is $\eta$, which as previously described we set by
fitting to the $P=0.53$ data. A close examination of FIG.
~\ref{difdw}(c) reveals that the P=0.72 data is not fit
quantitatively by either the finite surface tension or zero surface
tension theory. As previously discussed, we suspect that the central
region may not be superfluid.

As illustrated in FIG. ~\ref{up}, surface tension has almost no
effect on the axial density of the majority component.  The
smallness of the effect is to be expected because the discontinuity
in $n_\uparrow$ at the domain wall is much smaller than the
discontinuity in $n_\downarrow$. Alternate explanations of the
double-peaked axial density difference, such as anharmonicities
\cite{zwierlein}, would cause distortions in $n_\uparrow$ instead of
$n_{\downarrow}$ and are not completely consistent with the
experimental data \cite{replyR}.

We also calculated, but do not show here, density profiles for
parameters corresponding to the MIT experiments. We find that
surface tension has a negligible effect on the density profile,
consistent with the fact that $(\lambda/N)^{1/3}$ is 10 times
smaller than at Rice.

We wish to emphasize how surprises it is to see surface tension, a
phenomenon generally associated with liquid in a gas. This
observation opens the possibility of other surface tension related
effects in cold atoms. In particular the surface tension could have
a large effect on collective modes and expansion. We speculate that
surface tension should play a role in the physics of analogous
systems, such as nuclear matter at high densities and quark-gluon
plasmas.

This work was supported by NSF grant PHY-0456261, and by the Alfred
P. Sloan Foundation. We are grateful to R. Hulet and W. Li for very
enlightening discussions, critical comments, and for providing us
with their experimental data. We also thank M. Zwierlein for
providing us with their latest results \cite{ketterle2}, and for
insightful critical comments. We thank T.-L. Ho for critical
comments.

\appendix
%\section{Appendix}\label{ap}
\textbf{APPENDIX}: In this appendix we calculate the domain wall
energy at the superfluid-normal interface by applying a  gradient
expansion to mean field theory. The domain wall energy in this
approximation is given by,
$E(l)=\int^{l/2}_{-l/2}dx[\gamma\mid\partial_{x}\Delta\mid^2+E_{bcs}(\Delta,h,\mu)-E_{n}(h,\mu)]$,
where $l$, the size of the domain wall, will be determined
variationally, along with $\Delta(x)$, the superfluid order
parameter. The energies in the superfluid and normal phases are
given by \cite{old2},

\begin{eqnarray}
E_{bcs}(\Delta,h,\mu)=\frac{1}{2\pi^2}\int^{k_{+}}_{k_{-}} k^2dk
(-h+E_k)\\ \nonumber+\frac{1}{2\pi^2}\int
k^2dk\left[\epsilon_k-E_k+\frac{\Delta^2m}{\hbar^2k^2}\right]
%\\\nonumber
-\frac{\Delta^2m}{4\pi\hbar^2a_s}
\end{eqnarray}
\begin{eqnarray}\label{Nenergy}
E_n(h,\mu)=-\frac{1}{15\pi^2}\biggr(\frac{2m}{\hbar^2}\biggr)^{\frac{3}{2}}[(\mu+h)^{\frac{5}{2}}+(\mu-h)^{\frac{5}{2}}]
\end{eqnarray}
where $k_{\pm}(\vec{r})=(\pm\sqrt{h^2-\Delta^2}+\mu)^{1/2}$,
$E_k^2=\epsilon_k^2+\Delta^2$, $\epsilon_k=\hbar^2k^2/2m -\mu$ and,
$a_s$ is the s-wave scattering length.

In order to calculate the coefficient of the gradient term $\gamma$,
we begin with the action $S=S_0+S_{int}$, where the free fermions
action is $S_0=\sum_{\sigma} \int_0^{\beta}d\tau\int d^3\vec{r}
\psi_{r\sigma}^{\dagger}[\partial_{\tau}-\mu_{\sigma}+\hbar^2\nabla^2/2m]\psi_{r\sigma}$,
the interaction is $S_{int}=-U\int_0^{\beta}d\tau\int
d^3\vec{r}\psi_{r\uparrow}^{\dagger}\psi_{r\downarrow}^{\dagger}\psi_{r\downarrow}\psi_{r\uparrow}$,
the atomic Fermi fields are $\psi_{r\sigma}$, imaginary time is
$\tau$, the inverse temperature is $\beta=1/T$ and the attractive
interaction between Fermi atoms is $-U$ with $U\geq 0$. After the
usual Hubbard-Stratonovich decoupling of the interaction term
\cite{book} and integrating out the Fermi fields, the partition
function is written as $\emph{Z}=\int \emph{D}\Delta
\emph{D}\Delta^{\ast}\textit{exp}[-S_{eff}(\Delta,\Delta^{\ast})]$,
where
$S_{eff}(\Delta,\Delta^{\ast})=\sum_{q,n}[A(q,\omega_n)|\Delta(q)|^2+B(q,\omega)|\Delta(q)|^4+...]$
and $A(q,\omega_n)=(1/U-T\sum_n\int
d^3\vec{k}/(2\pi)^3G_{\uparrow}(k+q/2,\omega_m)G_{\downarrow}(-k+q/2,\omega_m+\omega_n))$
with
$G_{\sigma}(k,\omega)^{-1}=\imath\omega-\hbar^2k^2/2m+\mu_{\sigma}$.
We assume that the dominant momentum dependence comes from the term
which is lowest order in superfluid order parameter. We sum over
Matsubara frequencies, defining $A(q)=\sum_n A(q,\omega_n)$ with
$\omega_n=(2n+1)\pi T$. In order to suppress the ultraviolet
divergences in the theory, we regularize~\cite{car} the interaction
with the s-wave scattering length by
$1/U=m/4\pi\hbar^2a_s+d^3\vec{k}/(2\pi)^3m/\hbar^2k^2$. We then
expand $A(q)$ to second order in q and take the zero temperature
limit, finding $A(q)=m/(4\pi \hbar^2a_s)-m\sqrt{\mu}/(4\pi
\hbar^2)+mq^2/(32\pi\hbar^2\sqrt{\mu})+{\cal O}(q^4)$, which means
that $\gamma=m/(32\pi\hbar^2\sqrt{\mu})$.

\begin{figure}
\includegraphics[width=\columnwidth]{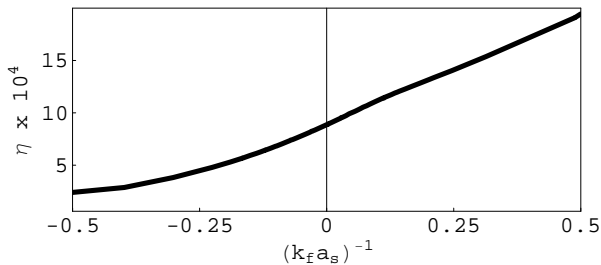}
\caption{The value of $\eta$ as a function of $(k_fa_s)^{-1}$ in
zero temperature BCS approximation. At unitarity $\eta=0.0009$ is
independent of the density. The Fermi wave vector $k_f$ is defined
as $k_f^3=3\pi^2n_{s}$.}\label{eta}
\end{figure}

Taking the ansatz $\Delta(x)=(\Delta_0/2)[\tanh(x/l)+1]$, where
$\Delta_0$ is the value of $\Delta$ on superfluid side of the domain
wall, we numerically minimize the surface energy $E(l)$ with respect
to $l$ to find the size of the domain wall $l_m$ and the domain wall
energy. We find $l_m<k_f^{-1}$, supporting our treatment of the
domain wall as very thin, but calling into question the validity of
our gradient expansion. Note that we do not expand $E_{bcs}$ in
powers of $\Delta$, but work with the exact expansion. Since we are
considering a domain wall between a region where $\Delta=0$ and
$\Delta\approx{\cal O}(E_f)$, any expansion in $\Delta$ would
require going to high order. To even capture the topology of the
free energy surface, one must expand to sixth order. Thus, previous
calculation of surface tension, such as Caldas's~\cite{caldas}
recent work, which are based on fourth order expansions are not
relevant to the physics described here. By repeating this
calculation at different $a_s$, and solving the BCS number equation
and gap equation~\cite{theja}, we find the quantity
$\eta=2mE(l_{m})/(\hbar^2n_{s}^{4/3})$ as a function of $a_s$ and
$n_{s}$, where $n_{s}$ is the density on the superfluid-normal
interface. In the limit of $a_s\rightarrow \infty$, we find
$\eta=0.9 \times 10^{-3}$, independent of the density and
polarization. However, as seen in FIG. ~\ref{eta}, $\eta$ has
density dependence away from unitarity. As $a_s\rightarrow 0^+$,
$\eta$ grows larger, hence the effects of surface tension is
stronger. Therefore, in the strong BEC regime, domain walls become
energetically prohibitive and the phase separated atomic system is
unstable against phase coexistence \cite{old2,theja,new1}. Recent
theoretical work by Imambekov \emph{et al} \cite{adilet} studied the
role of gradient terms in this deep BEC limit.

The value of $\eta$ obtained from fitting to experimental data
agrees well with our mean field calculation. We believe that this
agreement is coincidental as mean field theory is not expected to
work well at unitarity. We also note that the experiment is
performed slightly away from the resonance where the mean field
approximation predicts a weak density dependance of $\eta$
\cite{note2}.


\begin{references}
\bibitem{shells}The superfluid and normal regions may be further
subdivided, depending on interaction parameters. See \cite{theja}.
\bibitem{note1} Laplace's law gives, $\delta
P/P=2 \sigma/RP$, where $R$ is the radius of curvature of the domain
wall. Typical values from \cite{randy} are $\sigma/(\hbar^2/2m) \sim
5.8 \times 10^{-14}$ ($\mu$m)$^{-4}$, $R \sim 236$ $\mu$m, and
$P/(\hbar^2/2m) \sim 2.7 \times 10^{-13}$ ($\mu$m)$^{-5}$.
\bibitem{note2} The Rice experiment has a characteristic density
corresponding to $(k_fa_s)^{-1}\sim 0.09$.
\end{references}
\end{document}